\documentclass{appolb}
\usepackage{graphicx}

\def\cp{$CP$\/}
\def\cpv{$CPV$\/}

\def\ra{\!\rightarrow\!}
\def\dbar{\overline{D}{}^{\,0}}
\def\kbar{\overline{K}{}^{\,0}}

\def\dklnu{$D^0\ra K^+\ell^-\nu$}
\def\dkpi{$D^0\ra K^+\pi^-$}
\def\dkk{$D^0\ra K^+K^-$}
\def\dpipi{$D^0\ra\pi^+\pi^-$}

\def\dkspp{$D^0\ra K^0_S\,\pi^+\pi^-$}
\def\dppp{$D^0\ra \pi^0\,\pi^+\pi^-$}
\def\dkskk{$D^0\ra K^0_S\,K^+ K^-$}
\def\dkppp{$D^0\ra K^+\pi^-\pi^+\pi^-$}

\def\simge{\mathrel{%
   \rlap{\raise 0.511ex \hbox{$>$}}{\lower 0.511ex \hbox{$\sim$}}}}
\def\simle{\mathrel{
   \rlap{\raise 0.511ex \hbox{$<$}}{\lower 0.511ex \hbox{$\sim$}}}}

\begin{document}


\title{
\vspace*{-1.0in}
\flushright{UCHEP-18-02} \\
\vskip0.40in
Charm Physics: another route towards New Physics\thanks{Presented at 
the XXIV Cracow Epiphany Conference on Advances in Heavy Flavor Physics}
}

\author{A.~J.~Schwartz
\address{Physics Department, University of Cincinnati, 
Cincinnati, Ohio 45221 USA}
}

\maketitle

\begin{abstract}
We summarize recent results for charm physics. These results span
several categories: charm mixing, indirect (time-dependent) \cp\ 
violation, direct (time-integrated) \cp\ violation, $T$ violation, 
semileptonic and leptonic decays, and decays of charm baryons.
\end{abstract}

\section{Introduction}

Many new measurements of $D$ meson decays and charm baryon decays
have been performed by the Belle, BaBar, LHCb, and BESIII experiments.
Each experiment has unique advantages: Belle and BaBar produce boosted
charmed hadrons in a low-background $e^+e^-$ environment; LHCb produces
very large event samples due the large $c\bar{c}$ production cross section 
in hadron collisions; and BESIII produces $D\overline{D}$ meson pairs in
a quantum-correlated state at threshold with very little background.
Here we review recent results from all four experiments. The measurements 
can be grouped into six areas: measurements of charm mixing, 
indirect (time-dependent) \cp\ violation, 
direct (time-integrated) \cp\ violation, $T$ violation, 
semileptonic and leptonic decays, and charm baryon decays.
Our review highlights results that are sensitive to New Physics
and thus can constrain extensions to the Standard Model (SM).

\section{Mixing and indirect \cp\ violation}

Measurements of mixing and \cp\ violation require accurate flavor 
tagging and precise measurement of decay times. The former is usually 
achieved by reconstructing neutral $D$ mesons originating from 
$D^{*+}\ra D^0\pi^+$ and $D^{*-}\ra \dbar\pi^-$ 
decays;\footnote{Throughout this paper, charge-conjugate 
modes are implicitly included unless stated otherwise.}
the charge of the accompanying $\pi^\pm$ tags the flavor 
of the $D$. The latter is achieved by measuring the 
displacement $\vec{\ell}$ between the $D^{*+}$ and $D^0$ 
decay vertices and dividing by the $D^0$ momentum: 
$t = (\vec{\ell}\cdot\hat{p}^{}_D) (M^{}_{D^0}/p^{}_D)$, where 
$M^{}_{D^0}$ is the $D^0$ mass~\cite{pdg}.
The $D^{*+}$ vertex position is taken to be the intersection of
$\vec{p}^{}_D$ with the beam spot profile for $e^+e^-$ 
experiments, and at the primary interaction vertex for 
$\bar{p}p$ and $pp$ experiments.

\cp\ violation (\cpv) arises from interference between two or more decay 
amplitudes. When one of these amplitudes arises from mixing, then 
the resulting \cpv\ is called {\it indirect}. Otherwise, when no 
mixing is involved, the \cpv\ is called {\it direct}. Current 
measurements of charm mixing and {\it indirect\/} \cpv\ 
determine mixing parameters $x$, $y$, or 
$x'=x\cos\delta + y\sin\delta$, 
$y'=y\cos\delta - x\sin\delta$,
where $\delta$ is a strong phase;
\cpv\ parameters $|q/p|$ and ${\rm Arg}(q/p)\equiv\phi$; 
and ``mixed'' observables 
$y^{}_{CP}\approx y\cos\phi - (|q/p|-|p/q|)x\sin\phi\,/2$ and 
$A^{}_\Gamma\approx (|q/p|-|p/q|)y\cos\phi\,/2 - x\sin\phi$. 
A value $|q/p|\neq 1$ gives rise to \cpv\ in mixing, and 
a value $\phi\neq 0$ gives rise to \cpv\ resulting from 
interference between a mixed amplitude and a direct decay 
amplitude. For further details of these quantities, see the 
review by the Heavy Flavor Averaging Group (HFLAV)~\cite{hflav}.
Here we present recent measurements of $A^{}_\Gamma$, $x'^2$, $y'$,
and $|q/p|$ by LHCb~\cite{lhcb:a_gamma,lhcb:K+p_prompt,lhcb:K+p_fromB}, 
and results of a global fit for mixing and \cpv\ by HFLAV.

\subsection{LHCb measurements}

\begin{figure}[htb]
\begin{center}
\vbox{
\includegraphics[width=8.0cm]{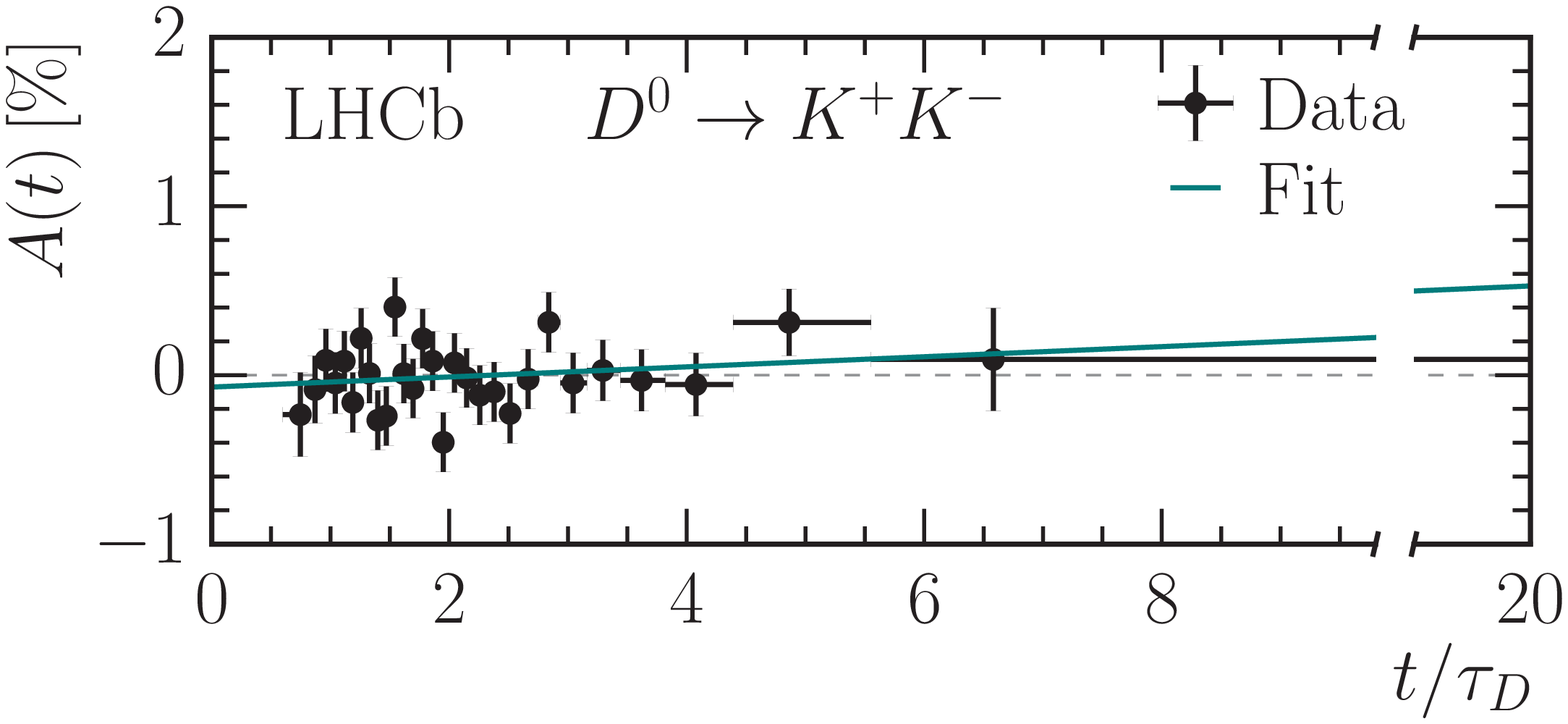}  \\
\includegraphics[width=8.0cm]{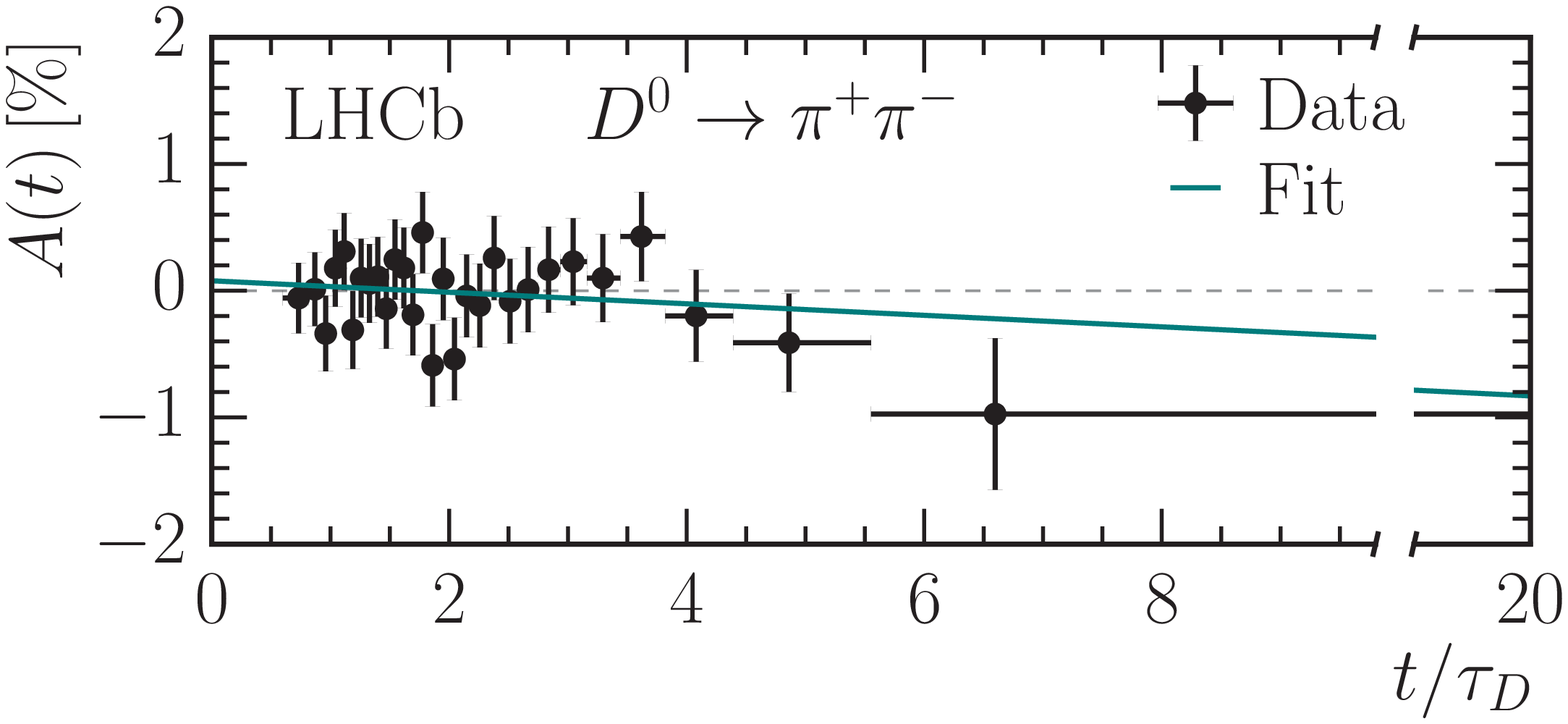} }
\end{center}
\vskip-0.20in
\caption{$A^{}_{CP}(t)$ for $D^0\ra K^+K^-$ (top) and 
$D^0\ra \pi^+\pi^-$ (bottom), from LHCb~\cite{lhcb:a_gamma}.}
\label{fig:1}
\end{figure}

LHCb recently measured $A^{}_\Gamma$ using their full Run I dataset 
of 3.0~fb$^{-1}$~\cite{lhcb:a_gamma}. This parameter is defined as
$A^{}_{\Gamma}\ \equiv\ 
(\hat{\tau}^{}_{\dbar\rightarrow f} - \hat{\tau}^{}_{D^0\rightarrow f})/
(\hat{\tau}^{}_{\dbar\rightarrow f} + \hat{\tau}^{}_{D^0\rightarrow f})$,
where $\hat{\tau}$ is the effective exponential lifetime of 
$D^0\ra f$ or $\dbar\ra f$ decays. This parameter can also be
measured via the time-dependent \cp\ asymmetry
\begin{equation}
A^{}_{CP}(t)\ \equiv\  
\frac{\displaystyle \Gamma(D^0(t)\ra f) - \Gamma(\dbar(t)\ra f)}
{\displaystyle \Gamma(D^0(t)\ra f) + \Gamma(\dbar(t)\ra f)} 
\ \approx\ a^f_{\rm direct} - A^{}_\Gamma \left(\frac{t}{\tau^{}_D}\right)\,,
\label{eqn:agamma}
\end{equation}
where $a^f_{\rm direct}$ represents the amount of direct \cpv\ in the 
decay, and $\tau^{}_D$ is the $D^0$ lifetime. This is the method used 
by LHCb. Here one fits for $D^0$ and $\dbar$ yields in bins of decay 
time and calculates the difference in yields over the sum; the resulting 
background-free distribution is fit to Eq.~(\ref{eqn:agamma}). The 
method is much less sensitive to the decay time resolution function,
which can be difficult to determine to high precision. The LHCb 
distribution is shown in Fig.~\ref{fig:1}. The fit results are
$A^{}_\Gamma(K^+K^-) = (-0.030\pm 0.032\pm0.010)\%$ and 
$A^{}_\Gamma(\pi^+\pi^-) = (0.046\pm 0.058\pm0.012)\%$, where
the first error is statistical and the second is systematic.
Combining these gives $A^{}_\Gamma = (-0.013\pm 0.028\pm0.010)\%$, 
which is the most precise result for $A^{}_\Gamma$ to-date.

LHCb also recently measured mixing parameters $x'^2$, $y'$, and 
the \cpv\ parameter $|q/p|$ using ``wrong-sign'' $D^0\ra K^+\pi^-$ 
decays and 3~fb$^{-1}$ of data~\cite{lhcb:K+p_prompt,lhcb:K+p_fromB}. 
Two separate analyses were performed, both using $D^{*+}\ra D^0\pi^+$ 
decays to tag the flavor of the $D^0$. However, the second analysis 
required the $D^*$ to originate from a $B\ra D^{*+}\mu^-\nu$ decay, 
and thus the $D^0$ flavor was also tagged by the $\mu^+$. The signal 
yield of the first analysis is 720\,000 events, while the yield for 
the ``double-tagged'' analysis is much less, 6680 events. However, 
upon combining the measurements, the latter adds $\sim\!10$\% in 
sensitivity due to very low background and increased acceptance 
at low $D^0$ decay times.

\begin{figure}[htb]
\begin{center}
\hbox{
\hskip-0.06in
\includegraphics[width=6.2cm]{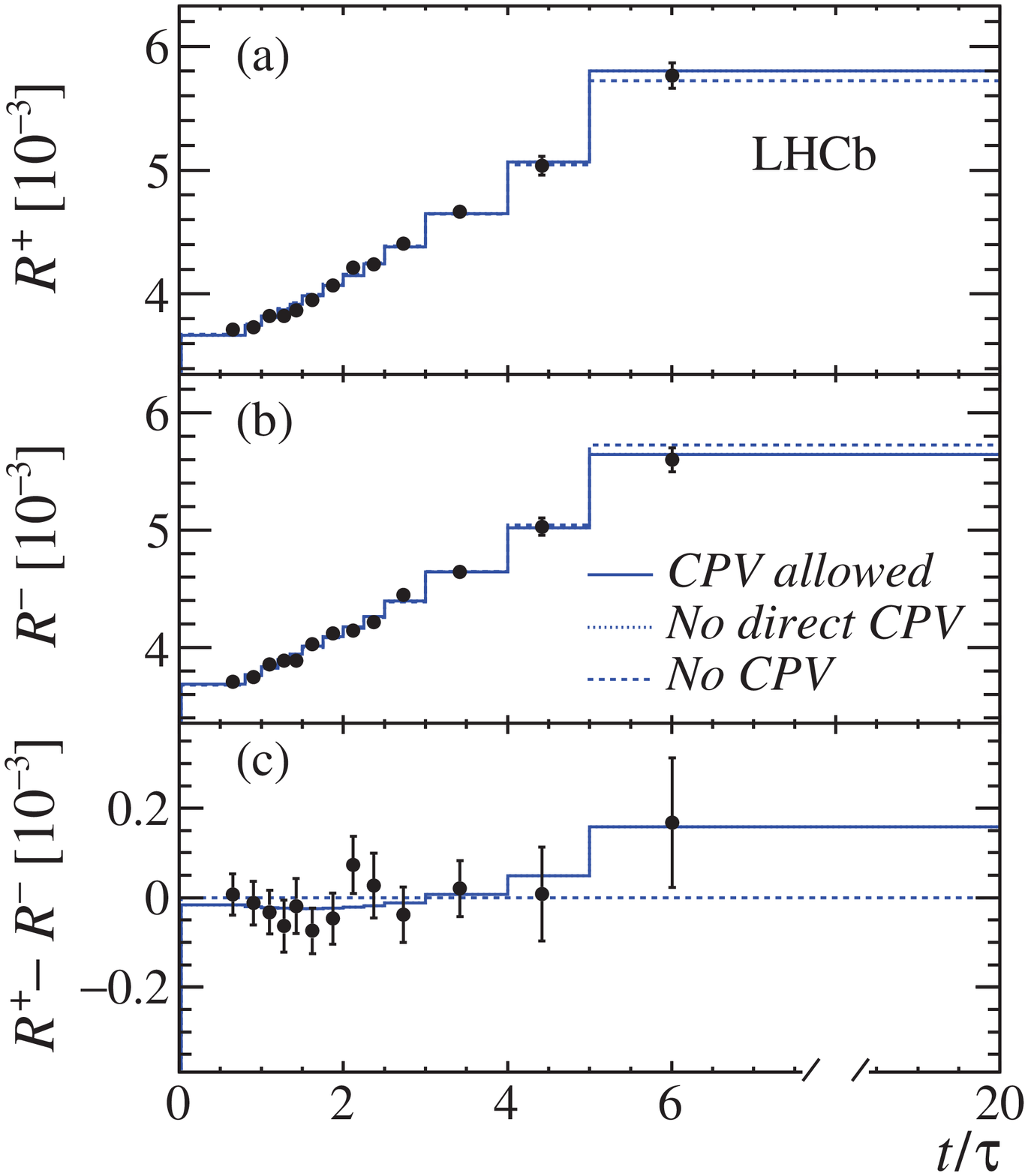}
\hskip0.10in
\includegraphics[width=6.2cm]{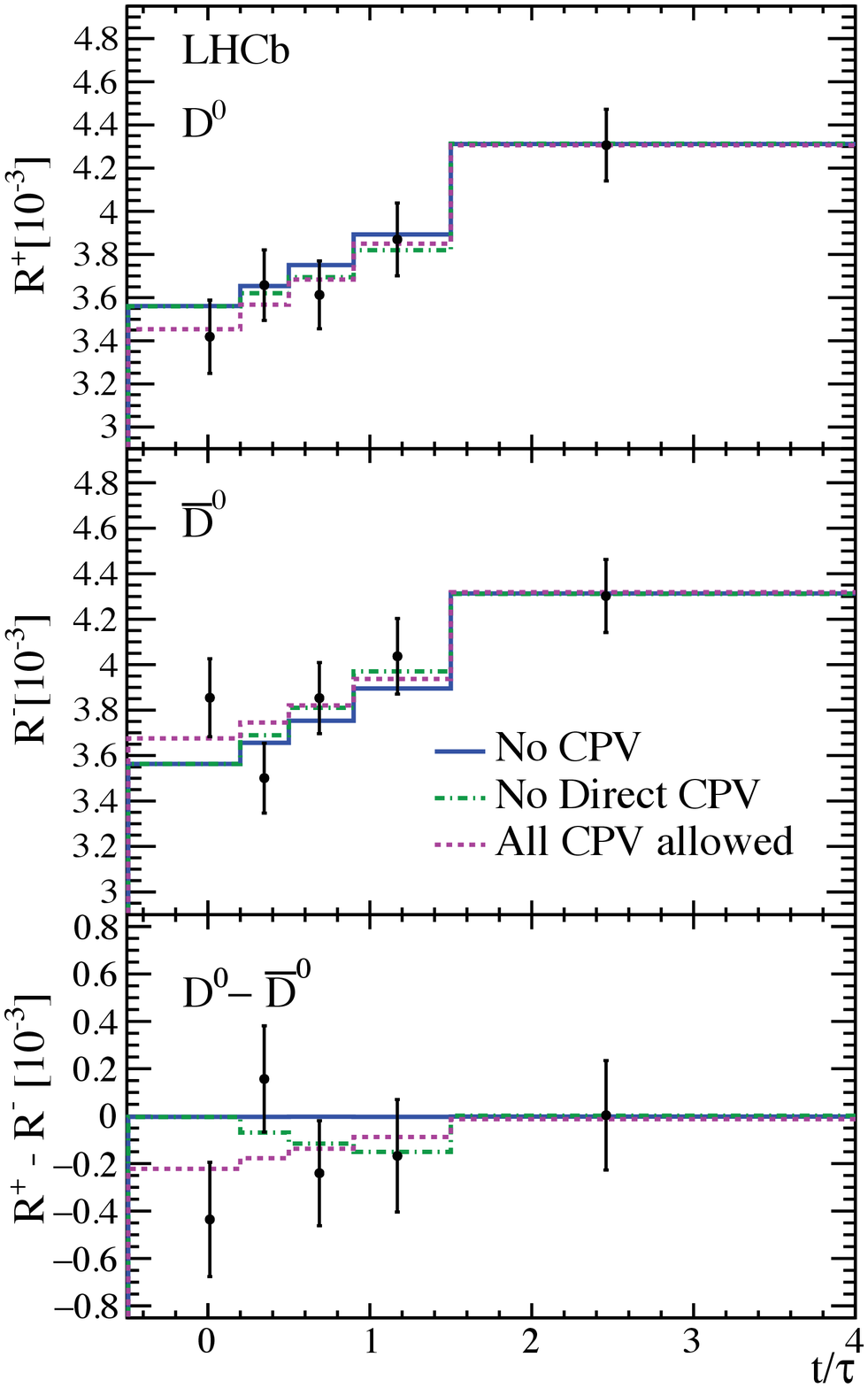} 
}
\end{center}
\vskip-0.30in
\caption{Ratios $R^+(t)$ and $R^-(t)$ for singly tagged (left) 
and doubly tagged (right) $D^0\ra K^+\pi^-$ decays, from 
LHCb~\cite{lhcb:K+p_prompt,lhcb:K+p_fromB}.}
\label{fig:2}
\end{figure}

The ratio of wrong-sign $D^0\ra K^+\pi^-$ decays to Cabibbo-favored 
``right-sign'' decays $D^0\ra K^-\pi^+$, and the ratio for 
$\dbar\ra K^-\pi^+$ to $\dbar\ra K^+\pi^-$, are respectively~\cite{Bergmann} 
\begin{eqnarray}
R^+(t) & = & 
R^{}_D \ +\  \left|\frac{q}{p}\right|\sqrt{R^{}_D} 
(y'\cos\phi - x'\sin\phi )(\overline{\Gamma}t) \ +\   
\left|\frac{q}{p}\right|^2 
\frac{(x'^2 + y'^2)}{4}(\overline{\Gamma}\,t)^2 \nonumber \\
\label{eqn:kp1} \\ 
R^-(t) & = & 
\overline{R}^{}_D \ +\  \left|\frac{p}{q}\right|\sqrt{\overline{R}^{}_D} 
y'\cos\phi + x'\sin\phi )(\overline{\Gamma}t) \ +\   
\left|\frac{p}{q}\right|^2 
\frac{(x'^2 + y'^2)}{4}(\overline{\Gamma}\,t)^2\,, \nonumber \\
\label{eqn:kp2}
\end{eqnarray}
where $R^{}_D$ is the ratio of amplitudes squared
$|{\cal A}(D^0\ra K^+\pi^-)|^2/|{\cal A}(D^0\ra K^-\pi^+)|^2$,
and 
$\overline{R}^{}_D = 
|{\cal A}(\dbar\ra K^-\pi^+)|^2/|{\cal A}(\dbar\ra K^+\pi^-)|^2$.
This measurement, like that for $A^{}_\Gamma$, is also performed 
in bins of decay time. For each bin, signal yields are obtained 
by fitting to variables $M^{}_D$ and $\Delta M = M^{}_{D^*} - M^{}_D$, 
and the ratios $R^+$ and $R^-$ calculated. The resulting 
(background-free) decay time distributions are shown in 
Fig.~\ref{fig:2}. Simultaneously fitting these distributions 
to Eqs.~(\ref{eqn:kp1}) and (\ref{eqn:kp2}) gives 
$x'^2 = (0.039\pm 0.023\pm0.014)\times 10^{-3}$ and
$y' = (0.528\pm 0.045\pm0.027)\%$. From the single-tagged 
analysis alone, a loose constraint $|q/p|\in [0.82,1.45]$ 
at 95\% CL is obtained.

\subsection{HFLAV global fit}

HFLAV calculates world average values of $A^{}_\Gamma$ and also $y^{}_{CP}$,
and inputs all $D^0$-$\dbar$ mixing measurements into a global 
fit to determine world average values for 10 parameters: 
$x$, $y$, $|q/p|$, $\phi$, $R^{}_D$,
direct \cpv\ parameters 
$A^{}_D$, $A^{}_K$, and $A^{}_\pi$, and
strong phase differences $\delta$ and $\delta^{}_{K\pi\pi}$.
The fit uses 49 observables from 
measurements of \dklnu, \dkk, \dpipi, \dkpi, 
$D^0\ra K^+\pi^-\pi^0$, 
\dkspp, \dppp, \dkskk, and \dkppp\ decays,
and double-tagged branching fractions measured at the 
$\psi(3770)$ resonance. Details are given in Ref.~\cite{hflav}.

The results of the fit are listed in Table~\ref{tab:hflav_results}.
Several fits are performed: 
({\it a}) assuming \cp\ conservation by fixing
$A^{}_D\!=\!0$, $A_K\!=\!0$, $A^{}_\pi\!=\!0$, $\phi\!=\!0$, 
and $|q/p|\!=\!1$;
({\it b}) assuming no direct \cpv\ in doubly Cabibbo-suppressed (DCS)
decays ($A^{}_D\!=\!0$);
({\it c}) assuming no direct \cpv\ in DCS decays and fitting 
for parameters $x^{}_{12}= 2|M^{}_{12}|/\Gamma$, 
$y^{}_{12}= \Gamma^{}_{12}/\Gamma$, and 
$\phi^{}_{12}= {\rm Arg}(M^{}_{12}/\Gamma^{}_{12})$,
where $M^{}_{12}$ and $\Gamma^{}_{12}$ are the off-diagonal
elements of the $D^0$-$\dbar$ mass and decay matrices, 
respectively; and
({\it d}) allowing full \cpv\ (floating all parameters). 

\begin{figure}[htb]
\begin{center}
\vskip0.20in
\hbox{
\includegraphics[width=5.9cm]{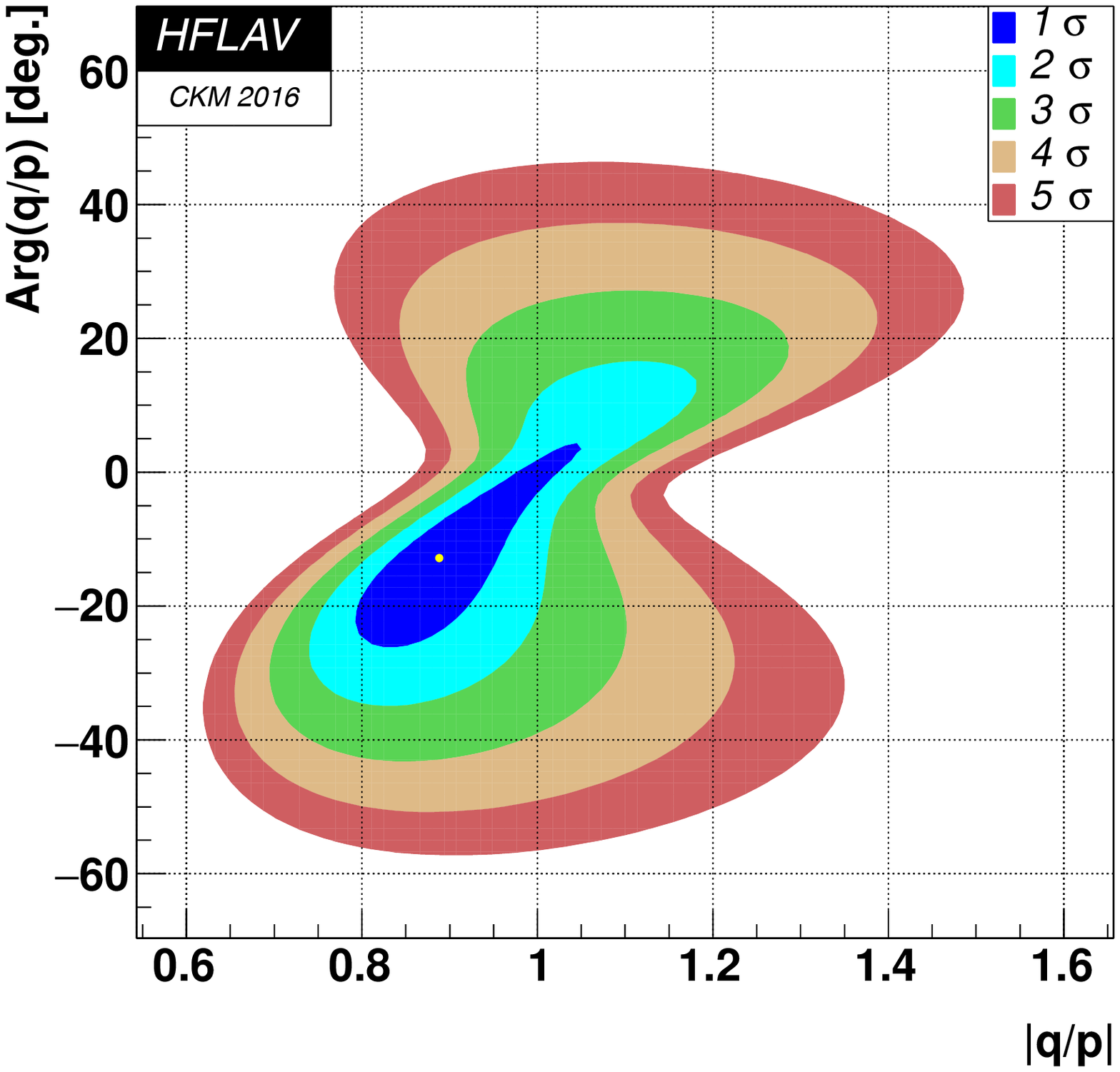}
\hskip0.10in
\includegraphics[width=5.9cm]{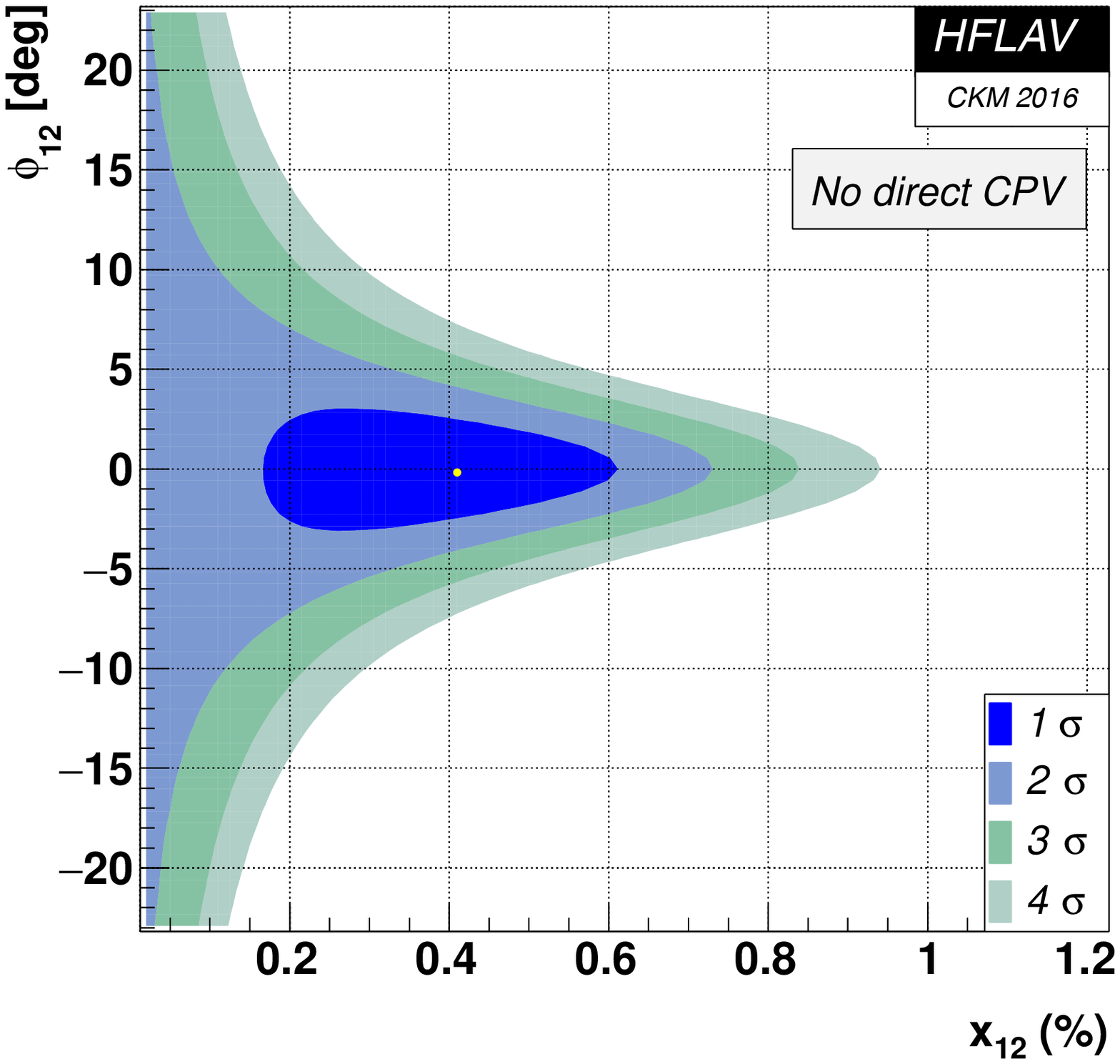} 
}
\end{center}
\vskip-0.20in
\caption{Confidence contours resulting from the HFLAV 
global fit to 49 observables, from Ref.~\cite{hflav}.}
\label{fig:hflav_contours}
\end{figure}

For fit {\it (b)}, in addition to $A^{}_D\!=\!0$ we impose the
constraint~\cite{Ciuchini,Kagan}
$\tan\phi = (1-|q/p|^2)/(1+|q/p|^2)\times (x/y)$, which reduces 
four independent parameters to three.\footnote{One can also use 
Eq.~(15) of Ref.~\cite{Grossman} to reduce four parameters to three.}
This constraint is imposed in two ways:
first floating $x$, $y$, and $\phi$ and from these deriving $|q/p|$; 
and alternatively floating $x$, $y$, and $|q/p|$ and from these 
deriving $\phi$. The central values obtained from the two fits 
are identical, but the first fit yields (MINOS) errors for $\phi$, 
while the second fit yields errors for $|q/p|$. For fit {\it (c)},
we float parameters $x^{}_{12}$, $y^{}_{12}$, and $\phi^{}_{12}$ 
and from these calculate~\cite{Kagan} $x$, $y$, $|q/p|$, and $\phi$;
these are then compared to measured values.
The $1\sigma\!-\!5\sigma$ contours in the two-dimensional parameter
spaces $(|q/p|, \phi)$ and $(x^{}_{12}, \phi^{}_{12})$ are shown in 
Fig.~\ref{fig:hflav_contours}. The HFLAV fit excludes the no-mixing 
point $x=y=0$ at $>11.5\sigma$, but the fit is consistent with 
\cp\ conservation ($|q/p| = 1$, $\phi = 0$).

\begin{table}
\begin{center}
\renewcommand{\arraystretch}{1.2}
\begin{tabular}{p{1.7cm} | p{2cm} p{2.6cm} p{2cm} p{2.6cm}}
\hline
Parameter & No \cpv & No direct \cpv & All \cpv & \cpv-allowed \\
 & & in DCS decays & allowed & (95\% CL) \\
\hline
$\begin{array}{c}
x\ (\%) \\ 
y\ (\%) \\ 
\delta^{}_{K\pi}\ (^\circ) \\ 
R^{}_D\ (\%) \\ 
A^{}_D\ (\%) \\ 
|q/p| \\ 
\phi\ (^\circ) \\
\delta^{}_{K\pi\pi}\ (^\circ)  \\
A^{}_{\pi} (\%) \\
A^{}_K (\%) \\
x^{}_{12}\ (\%) \\ 
y^{}_{12}\ (\%) \\ 
\phi^{}_{12} (^\circ)
\end{array}$ 
& 
$\begin{array}{c}
0.46\,^{+0.14}_{-0.15} \\
0.62\,\pm 0.08 \\
8.0\,^{+9.7}_{-11.2} \\
0.348\,^{+0.004}_{-0.003} \\
- \\
- \\
- \\
20.4\,^{+23.3}_{-23.8} \\
- \\
- \\
- \\
- \\
- 
\end{array}$ 
&
$\begin{array}{c}
0.41\,^{+0.14}_{-0.15}\\
0.61\,\,\pm 0.07 \\
4.8\,^{+10.4}_{-12.3}\\
0.347\,\,^{+0.004}_{-0.003} \\
- \\
0.999\,\pm 0.014 \\
0.05\,\,^{+0.54}_{-0.53} \\ 
22.6\,^{+24.1}_{-24.4} \\ 
0.02\,\pm 0.13 \\
-0.11\,\pm 0.13 \\
0.41\,^{+0.14}_{-0.15}\\
0.61\,\pm 0.07 \\
-0.17\,\pm 1.8 
\end{array}$ 
&
$\begin{array}{c}
0.32\,\pm 0.14 \\
0.69\,\,^{+0.06}_{-0.07}\\
15.2\,^{+7.6}_{-10.0} \\
0.349\,^{+0.004}_{-0.003} \\
-0.88\,\pm 0.99 \\
0.89\,^{+0.08}_{-0.07} \\ 
-12.9\,^{+9.9}_{-8.7} \\ 
31.7\,^{+23.5}_{-24.2} \\
0.01\,\pm 0.14 \\
-0.11\,\pm 0.13 \\
 \\
 \\
 \\
\end{array}$ 
&
$\begin{array}{c}
\left[ 0.04,\, 0.62\right] \\
\left[ 0.50,\, 0.80\right] \\
\left[ -16.8,\, 30.1\right] \\
\left[ 0.342,\, 0.356\right] \\
\left[ -2.8,\, 1.0\right] \\
\left[ 0.77,\, 1.12\right] \\\
\left[ -30.2,\, 10.6\right] \\
\left[ -16.4,\, 77.7\right] \\
\left[ -0.25,\, 0.28\right] \\
\left[ -0.37,\, 0.14\right] \\
\left[ 0.10,\, 0.67\right] \\
\left[ 0.47,\, 0.75\right] \\
\left[ -5.3,\, 4.4\right] \\
\end{array}$ 
\\
\hline
\end{tabular}
\end{center}
\vskip-0.10in
\caption{Results of the HFLAV global fit to 49 observables, from Ref.~\cite{hflav}.}
\label{tab:hflav_results}
\end{table}

\section{Direct \cp\ violation}

In addition to searches for indirect \cpv\ in $D$ decays, there 
have been many searches for direct \cpv. Such searches consist of 
time-integrated measurements, i.e., they do not require measuring 
decay times. However, flavor-tagging is important and, for high 
precision measurements, usually needs to be corrected for small
systematic effects such as a possible charge asymmetry in the 
reconstruction of the low momentum $\pi^\pm$ originating from 
$D^{*\pm}\ra D\pi^\pm$ 
decays. The results to-date are listed in 
Table~\ref{tab:dcpv_dzero} for $D^0$ decays, 
Table~\ref{tab:dcpv_dplus} for $D^+$ decays, and
Table~\ref{tab:dcpv_ds} for $D^+_s$ decays. 
There are recent results from Belle 
($D^0\ra K^0_SK^0_S$~\cite{belle:dcpv_KSKS}, 
$D^0\ra \rho^0/\phi/\overline{K}^{*0}\gamma$~\cite{belle:dcpv_r0g}, 
$D^+\ra\pi^+\pi^0$~\cite{belle:dcpv+_pp0}) 
and LHCb 
($D^0\ra\pi^+\pi^-$~\cite{lhcb:dcpv_KK},
$D^0\ra K^+K^-$~\cite{lhcb:dcpv_KK}, 
$D^0\ra\pi^+\pi^-\pi^+\pi^-$~\cite{lhcb:dcpv_pppp},
$D^+_{(s)}\ra\eta'\pi^+$~\cite{lhcb:dcpv+_etapp}).
In all cases the results are consistent with no~\cpv. 
Several measurements have a precision of 0.2\% or smaller.

\begin{table}
\begin{center}
\begin{tabular}{|l|l|c|l|}
\hline
Decay & Channel & World avg. or    & Most precise  \\
      &         & most precise (\%) & measurement   \\
\hline
\hskip-0.10in
\begin{tabular}{l}
Cabibbo- \\
favored
\end{tabular} 
& $D^0\ra K^-\pi^+$    & $0.3\,\pm 0.7$ & CLEO 2014~\cite{cleo:dcpv_K-p} \\
& $D^0\ra K^0_S \pi^0$ & $-0.20\,\pm 0.17$ & Belle 2014~\cite{belle:dcpv_h0p0} \\
& $D^0\ra K^-\pi^+\pi^0$ & $0.1\,\pm 0.5$ & CLEO 2014~\cite{cleo:dcpv_K-pp0} \\
& $D^0\ra K^0_S\pi^+\pi^-$ & $-0.08\pm 0.77$ & CDF 2012~\cite{cdf:dcpv_KSpp} \\
& $D^0\ra K^-\pi^+ \pi^-\pi^+$ & $0.2\,\pm 0.5$ & CLEO 2014~\cite{cleo:dcpv_K-ppp} \\
& $D^0\ra \eta\,K^0_S$ & $0.54\,\pm 0.53$ & Belle 2011~\cite{belle:dcpv_KSeta} \\
& $D^0\ra \eta' K^0_S$ & $0.98\,\pm 0.68$ & Belle 2011~\cite{belle:dcpv_KSeta} \\
\hline
\hskip-0.10in
\begin{tabular}{l}
Singly \\
Cabibbo- \\
suppressed
\end{tabular} 
& $D^0\ra \pi^+ \pi^-$ & $0.00\,\pm 0.15$ & LHCb 2017~\cite{lhcb:dcpv_KK}  \\
& $D^0\ra \pi^0 \pi^0$ & $-0.03\,\pm 0.64$ & Belle 2014~\cite{belle:dcpv_h0p0} \\
& $D^0\ra \pi^+\pi^-\pi^0$ & $0.32\,\pm 0.42$ & $\Biggl\{$\hskip-0.05in
\begin{tabular}{l} 
LHCb 2015~\cite{lhcb:dcpv_ppp0} \\
BaBar 2008~\cite{babar:dcpv_KKp0} 
\end{tabular} \\
& $D^0\ra K^0_S K^0_S$ & $-0.02\,\pm 1.54$ & Belle 2017~\cite{belle:dcpv_KSKS}  \\
& $D^0\ra K^+K^- $    & $-0.16\,\pm 0.12$ & LHCb 2017~\cite{lhcb:dcpv_KK}  \\
& $D^0\ra K^+ K^-\pi^0$ & $-1.00\,\pm 1.69$ & BaBar 2008~\cite{babar:dcpv_KKp0} \\
& $D^0\ra K^0_S\,K^\pm\pi^+ $  & - & LHCb 2016~\cite{lhcb:dcpv_KSKp}  \\
& $D^0\ra \pi^+\pi^-\pi^+\pi^-$ & - & LHCb 2017~\cite{lhcb:dcpv_pppp}  \\
& $D^0\ra K^+K^-\pi^+\pi^-$     & - & LHCb 2013~\cite{lhcb:dcpv_KKpp}  \\
\hline
\hskip-0.10in
\begin{tabular}{l}
Doubly \\
Cabibbo- \\
suppressed
\end{tabular} 
& $D^0\ra K^+\pi^-\pi^0$ & $-0.14\,\pm 5.17 $ & Belle 2005~\cite{belle:dcpv_K+pp0} \\
& $D^0\ra K^+\pi^- \pi^+\pi^-$ & $-1.8\,\pm 4.4$ & Belle 2005~\cite{belle:dcpv_K+pp0} \\
\hline
\hskip-0.10in
\begin{tabular}{l}
Radiative 
\end{tabular}
& $D^0\ra \rho^0\gamma$ & $5.6\,\pm 15.2$ & Belle 2017~\cite{belle:dcpv_r0g}  \\
& $D^0\ra \phi\gamma$ & $-9.4\,\pm 6.6$ & Belle 2017~\cite{belle:dcpv_r0g} \\
& $D^0\ra \overline{K}^{\,*0}\gamma$ & $-0.3\,\pm 2.0$ & Belle 2017~\cite{belle:dcpv_r0g}  \\
\hline
\end{tabular}
\end{center}
\vskip-0.10in
\caption{Time-integrated \cp\ asymmetries for hadronic $D^0$ decays. 
The world averages are from HFLAV~\cite{hflav_web}.}
\label{tab:dcpv_dzero}
\end{table}

\begin{table}
\begin{center}
\begin{tabular}{|l|l|c|l|}
\hline
Decay & Channel & World avg. or    & Most precise  \\
      &         & most precise (\%) & measurement   \\
\hline
\hskip-0.10in
\begin{tabular}{l}
Cabibbo- \\
favored
\end{tabular} 
& $D^+\ra K^0_S\,\pi^+$    & $-0.41\,\pm 0.09$ & Belle 2012~\cite{belle:dcpv+_KSp} \\
& $D^+\ra K^0_S\,\pi^+\pi^0$    & $-0.1\,\pm 0.7$ & CLEO 2014~\cite{cleo:dcpv+_KSpp0} \\
& $D^+\ra K^0_S\,\pi^+\pi^+\pi^-$  & $0.0\,\pm 1.2$ & CLEO 2014~\cite{cleo:dcpv+_KSpp0} \\
& $D^+\ra K^-\pi^+\pi^+$    & $-0.18\,\pm 0.16$ & D\O\ 2014~\cite{d0:dcpv+_Kpp} \\
& $D^+\ra K^-\pi^+\pi^+\pi^0$    & $-0.3\,\pm 0.7$ & CLEO 2014~\cite{cleo:dcpv+_KSpp0} \\
\hline
\hskip-0.10in
\begin{tabular}{l}
Singly \\
Cabibbo- \\
suppressed
\end{tabular} 
& $D^+\ra \pi^+\pi^0$    & $2.3\,\pm 1.3$ & Belle 2017~\cite{belle:dcpv+_pp0}  \\
& $D^+\ra \pi^+\pi^+\pi^-$  & - & LHCb 2014~\cite{lhcb:dcpv+_ppp}  \\
& $D^+\ra K^0_S K^+$    & $0.11\,\pm 0.17$ & LHCb 2014~\cite{lhcb:dcpv+_KSK} \\
& $D^+\ra K^0_S K^+\pi^+\pi^-$ & $-4.2\,\pm 6.8$ & FOCUS 2005~\cite{focus:dcpv+_KSKpp} \\
& $D^+\ra K^+ K^-\pi^+$    & $0.32\,\pm 0.31$ & BaBar 2013~\cite{babar:dcpv+_KKp} \\
& $D^+\ra \eta\,\pi^+$    & $1.0\,\pm 1.0$ & Belle 2011~\cite{belle:dcpv+_etap} \\
& $D^+\ra \eta'\pi^+$    & $-0.61\,\pm 0.90$ & LHCb 2017~\cite{lhcb:dcpv+_etapp}  \\
\hline
\hskip-0.10in
\begin{tabular}{l}
Doubly \\
Cabibbo- \\
suppressed
\end{tabular} 
& $D^+\ra K^+\pi^0$    & $-3.5\,\pm 10.7$ & CLEO 2010~\cite{cleo:dcpv+_K+p0} \\
\hline
\end{tabular}
\end{center}
\vskip-0.10in
\caption{Time-integrated \cp\ asymmetries for hadronic $D^+$ decays.
The world averages are from HFLAV~\cite{hflav_web}.}
\label{tab:dcpv_dplus}
\end{table}

\begin{table}
\begin{center}
\begin{tabular}{|l|l|c|l|}
\hline
Decay & Channel & World avg. or    & Most precise  \\
      &         & most precise (\%) & measurement   \\
\hline
\hskip-0.10in
\begin{tabular}{l}
Cabibbo- \\
favored
\end{tabular} 
& $D^+_s\ra K^0_S\,K^+$    & $0.08\,\pm 0.26$ & BaBar 2013~\cite{babar:dcpvs_KSK} \\
& $D^+_s\ra K^0_S\,K^+\pi^0$    & $-1.6\,\pm 6.1$ & CLEO 2013~\cite{cleo:dcpvs_KKp} \\
& $D^+_s\ra K^0_S\,K^0_S\,\pi^+$    & $3.1\,\pm 5.2$ & CLEO 2013~\cite{cleo:dcpvs_KKp} \\
& $D^+_s\ra K^+ K^-\pi^+$    & $-0.5\,\pm 0.9$ & CLEO 2013~\cite{cleo:dcpvs_KKp} \\
& $D^+_s\ra K^+ K^-\pi^+\pi^0$    & $0.0\,\pm 3.0$ & CLEO 2013~\cite{cleo:dcpvs_KKp} \\
& $D^+_s\ra K^- K^0_S\,\pi^+\pi^+$    & $4.1\,\pm 2.8$ & CLEO 2013~\cite{cleo:dcpvs_KKp} \\
& $D^+_s\ra K^0_S\,K^+\pi^+\pi^-$    & $-5.7\,\pm 5.4$ & CLEO 2013~\cite{cleo:dcpvs_KKp} \\
& $D^+_s\ra \eta\,\pi^+$        & $1.1\,\pm 3.1$ & CLEO 2013~\cite{cleo:dcpvs_KKp} \\
& $D^+_s\ra \eta\,\pi^+\pi^0$    & $-0.5\,\pm 4.4$ & CLEO 2013~\cite{cleo:dcpvs_KKp} \\
& $D^+_s\ra \eta'\pi^+$    & $-0.82\,\pm 0.50$ & LHCb 2017~\cite{lhcb:dcpv+_etapp}  \\
& $D^+_s\ra \eta'\pi^+\pi^0$    & $-0.4\,\pm 7.6$ & CLEO 2013~\cite{cleo:dcpvs_KKp} \\
\hline
\hskip-0.10in
\begin{tabular}{l}
Singly \\
Cabibbo- \\
suppressed
\end{tabular} 
& $D^+_s\ra K^0_S\,\pi^+$    & $0.38\,\pm 0.49$ & LHCb 2014~\cite{lhcb:dcpvs_KSp} \\
& $D^+_s\ra K^+\pi^0$    & $-27\,\pm 24$ & CLEO 2010~\cite{cleo:dcpv+_K+p0} \\
& $D^+_s\ra K^+\pi^+\pi^-$    & $4.5\,\pm 4.8$ & CLEO 2013~\cite{cleo:dcpvs_KKp} \\
& $D^+_s\ra \eta\,K^+$    & $9.3\,\pm 15.2$ & CLEO 2010~\cite{cleo:dcpv+_K+p0} \\
& $D^+_s\ra \eta' K^+$    & $6\,\pm 19$ & CLEO 2010~\cite{cleo:dcpv+_K+p0} \\
\hline
\hskip-0.10in
\begin{tabular}{l}
Annihilation
\end{tabular} 
& $D^+_s\ra \pi^+\pi^-\pi^+$    & $-0.7\,\pm 3.1$ & CLEO 2013~\cite{cleo:dcpvs_KKp} \\
\hline
\end{tabular}
\end{center}
\vskip-0.10in
\caption{Time-integrated \cp\ asymmetries for hadronic $D^+_s$ decays.
The world averages are from HFLAV~\cite{hflav_web}.}
\label{tab:dcpv_ds}
\end{table}

\section{$T$ violation}

\begin{figure}[htb]
\begin{center}
\includegraphics[width=3.4cm,angle=90.]{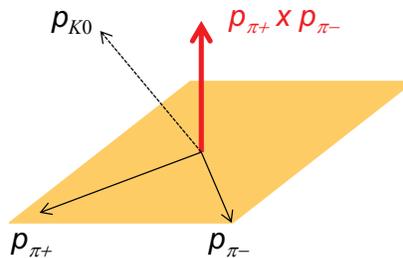}
\end{center}
\vskip-0.10in
\caption{Decay topology for $D^0\ra K^0_S\,\pi^+\pi^-\pi^0$. }
\label{fig:Tviolation}
\end{figure}

Belle recently measured the $T$-violating parameter $a^{}_T$ 
for Cabbibo-favored $D^0\ra K^0_S\,\pi^+\pi^-\pi^0$ decays using
their full dataset of 966~fb$^{-1}$~\cite{belle:Tviol_KSppp0}. 
The method used is similar to that used for earlier measurements 
of $D^0\ra K^+ K^-\pi^+\pi^-$ decays (BaBar~\cite{babar:Tviol_KKpp}, 
LHCb~\cite{lhcb:Tviol_KKpp}) and $D^+_{(s)}\ra K^+ K^0_S\,\pi^+\pi^-$ 
decays (BaBar~\cite{babar:Tviol_KSKpp}). This method is as follows. 
From the momenta of the daughter particles, one calculates the 
$T$-odd quantities
\begin{eqnarray}
C^{}_T & \equiv & \vec{p}^{}_{K_S}
\cdot \left( \vec{p}^{}_{\pi^+} \times \vec{p}^{}_{\pi^-} \right) 
\end{eqnarray}
for $D^0\ra K^0_S\,\pi^+\pi^-\pi^0$ decays, and
\begin{eqnarray}
\overline{C}^{}_T & \equiv & \vec{p}^{}_{K_S}
\cdot \left( \vec{p}^{}_{\pi^-} \times \vec{p}^{}_{\pi^+} \right)
\end{eqnarray}
for $\dbar\ra K^0_S\,\pi^-\pi^+\pi^0$ decays. One integrates
these quantities to construct the $T$-odd observables
\begin{eqnarray}
A^{}_T & \equiv & 
\frac{\displaystyle \Gamma(C^{}_T >0) - \Gamma(C^{}_T <0)}{\Gamma}
\end{eqnarray}
for $D^0$ decays, and 
\begin{eqnarray}
\bar{A}^{}_T & \equiv & 
\frac{\displaystyle \overline{\Gamma}(-\overline{C}^{}_T >0) - 
\overline{\Gamma}(-\overline{C}^{}_T <0)}{\overline{\Gamma}}
\end{eqnarray}
for $\dbar$ decays. As illustrated in Fig.~\ref{fig:Tviolation}, 
these observables correspond to the difference between the $K^0_S$ 
momentum projecting above the $(\pi^+, \pi^-)$ decay plane, and 
the momentum projecting below. Both $A^{}_T$ and $\bar{A}^{}_T$ 
may be nonzero due to either interference
between strong phases in the decay amplitude, or $T$ violation.
A difference due to strong phases would be the same for $A^{}_T$ 
and $\bar{A}^{}_T$, and thus the difference 
$a^{}_T \equiv (A^{}_T - \bar{A}^{}_T)/2$ isolates 
the $T$-violating effect~\cite{BensalemLondon}. 
(This asymmetry is also \cp-violating, so $CPT$ 
conservation implies $T$ violation.)

The Belle measurement for $D^0\ra K^0_S\,\pi^+\pi^-\pi^0$
has good precision, as the signal yield is large and 
backgrounds are low. Belle fits for the signal yields 
of four independent subsamples:
$\{D^0,C^{}_T\!>\!0\}$, 
$\{D^0,C^{}_T\!<\!0\}$, 
$\{\dbar,C^{}_T\!>\!0\}$, and 
$\{\dbar,C^{}_T\!<0\!\}$.
The resulting yields give $a^{}_T = (-0.028\,\pm 0.138\,^{+0.023}_{-0.076})\%$,
which is consistent with zero. As the four-body final state results mainly 
from two- and three-body intermediate states, Belle also divides the event 
sample into 
ranges of $M(\pi^+\pi^-\pi^0)$, $M(\pi^\pm\pi^0)$, and $M(K^0_S\pi^\pm)$ 
invariant masses to isolate $K^0_S\,\omega$, $K^0_S\,\eta$, $K^{*-}\rho^+$, 
$K^{*+}\rho^-$, $K^{*-}\pi^+\pi^0$, $K^{*+}\pi^-\pi^0$, $K^{*0}\pi^+\pi^-$, 
and $K^0_S\,\rho^+\pi^-$ intermediate states. For these subsamples, 
$a^{}_T$ is recalculated. The results are listed in 
Table~\ref{tab:Tviolation} and are all consistent 
with zero, i.e., no $T$ violation is seen. Previous 
measurements of $a^{}_T$ for 
$D^0\ra K^+K^-\pi^+\pi^-$~\cite{babar:Tviol_KKpp,lhcb:Tviol_KKpp}
and $D^+_{(s)}\ra K^+ K^0_S\,\pi^+\pi^-$~\cite{babar:Tviol_KSKpp} 
also show no evidence for $T$ violation.

\begin{table}[htb]
\centering
\renewcommand{\arraystretch}{1.2}
\begin{tabular}{l|ccc}
\hline \hline
Resonance & Invariant mass & $A^{}_T$ ($\times 10^{-2}$) & 
$a^{}_T$ ($\times 10^{-3}$) \\
    & range (GeV/$c^{2}$)    &    &          \\
\hline
$K^{0}_S\,\omega$	& $0.762<M_{\pi^{+}\pi^{-}\pi^{0}}<0.802$ & 
   $3.6\pm 0.5\pm 0.5$ & $-1.7 \pm 3.2 \pm 0.7$	\\
$K^{0}_S\,\eta$	& $M_{\pi^{+}\pi^{-}\pi^{0}} < 0.590$ & 
   $0.2\pm 1.3\pm 0.4$ & $4.6 \pm 9.5  \pm 0.2$	\\
$K^{*-}\rho^{+}$ & $0.790 <M_{K_{S}^{0}\pi^{-}}  <0.994$ & 
   $6.9\pm 0.3\,^{+0.6}_{-0.5}$ & $0.0 \pm 2.0\,^{+1.6}_{-1.4}$ \\
  & $0.610 <  M_{\pi^{+}\pi^{0}}~<0.960$	&		&    \\
$K^{*+}\rho^{-}$ & $0.790 < M_{K_{S}^{0}\pi^{+}} < 0.994$ & 
   $22.0\pm 0.6\pm 0.6$   & $1.2 \pm 4.4\,^{+0.3}_{-0.4}$ \\
  & $0.610 <  M_{\pi^{-}\pi^{0}}~<0.960$	&	&        \\
$K^{*-}\pi^{+}\pi^{0}$ & $0.790 < M_{K_{S}^{0}\pi^{-}} < 0.994$ &  
   $25.5\pm 0.7\pm 0.5$ & $-7.1 \pm 5.2\,^{+1.2}_{-1.3}$  \\
$K^{*+}\pi^{-}\pi^{0}$ & $0.790 < M_{K_{S}^{0}\pi^{+}} < 0.994$ & 
   $24.5\pm 1.0\,^{+0.7}_{-0.6}$  & $-3.9 \pm 7.3\,^{+2.4}_{-1.2}$ \\
$K^{*0}\pi^{+}\pi^{-}$ & $0.790 <M_{K_{S}^{0}\pi^{0}}  <0.994$ &  
   $19.7\pm 0.8\,^{+0.4}_{-0.5}$  & $0.0 \pm 5.6\,^{+1.1}_{-0.9}$  \\
$K^{0}_S\,\rho^{+}\pi^{-}$  & $0.610 < M_{\pi^{+}\pi^{0}} < 0.960$ & 
   $13.2\pm 0.9\pm 0.4$   & $7.6 \pm 6.1\,^{+0.2}_{-0.0}$  \\
Rest  	& $-$ & $20.5\pm 1.0\,^{+0.5}_{-0.6}$ & $1.8 \pm 7.4\,^{+2.1}_{-5.3}$ \\
\hline \hline
\end{tabular}
\caption{Values of $A^{}_T$ and $a^{}_T$ for different 
regions of $D^0\ra K_S^0\,\pi^+\pi^-\pi^0$ phase space, 
from Belle~\cite{belle:Tviol_KSppp0}. $M_{ij[k]}$ indicates 
the invariant mass of mesons $i$ and $j$ [and $k$].}
\label{tab:Tviolation}
\end{table}

\section{Semileptonic and leptonic decays}

Semileptonic and leptonic $D$ decays are easier to understand 
theoretically than hadronic decays. Their decay rates are 
parameterized as 
\begin{eqnarray*}
\frac{d\Gamma(D\ra P\ell^+\nu)}{dq^2} & = & 
\frac{G^2_F}{24\pi^3} |f^+(q^2)|^2 |V^{}_{cs,cd}|^2 p^{*3} 
\label{eqn:semileptonic} 
\end{eqnarray*}
and
\begin{eqnarray*}
\Gamma(D^+_{(s)}\ra\ell^+\nu) & = & \frac{G^2_F}{8\pi} 
f^2_{D_{(s)}} |V^{}_{cs,cd}|^2 m^{}_{D}\,m^2_\ell 
\left(1-\frac{m^2_\ell}{m^2_{D}}\right)^2\,,
\label{eqn:leptonic} 
\end{eqnarray*}
where $V^{}_{cs}$ and $V^{}_{cd}$ are CKM matrix elements, 
$p^*$ is the magnitude of the momentum of the final 
state hadron in the $D$ rest frame,
$f^+(q^2)$ is a form factor evaluated at
$q^2=(P^{}_D - P^{}_P)^2 = (P^{}_\ell + P^{}_\nu)^2$, 
and $f^{}_{D_{(s)}}$ is the $D^+_{(s)}$ decay constant.
Thus, with knowledge of $f^+(q^2)$ or $f^{}_{D_{(s)}}$
(e.g., from lattice QCD calculations), semileptonic and
leptonic decay rates determine $|V^{}_{cd}|$ and $|V^{}_{cs}|$.
Alternatively, assuming values of $|V^{}_{cd}|$ and 
$|V^{}_{cs}|$ (e.g., from CKM unitarity), the decay  
rates determine $f^+(q^2)$ and $f^{}_{D_{(s)}}$. These 
form factor and decay constant values can be compared 
to theory predictions.

\subsection{BESIII results}

\begin{figure}[htb]
\begin{center}
\hbox{
\includegraphics[width=6.3cm]{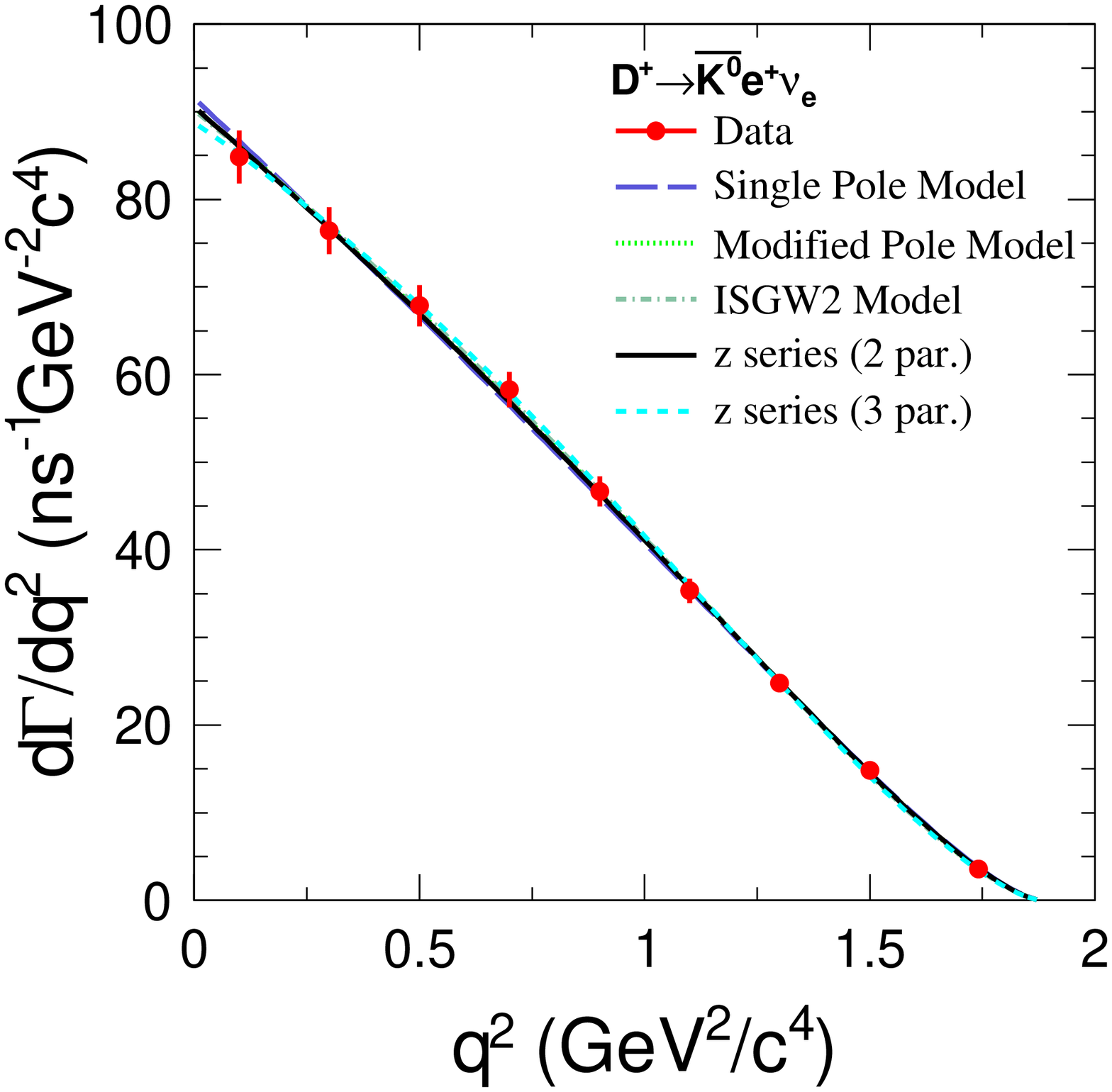}
\includegraphics[width=6.3cm]{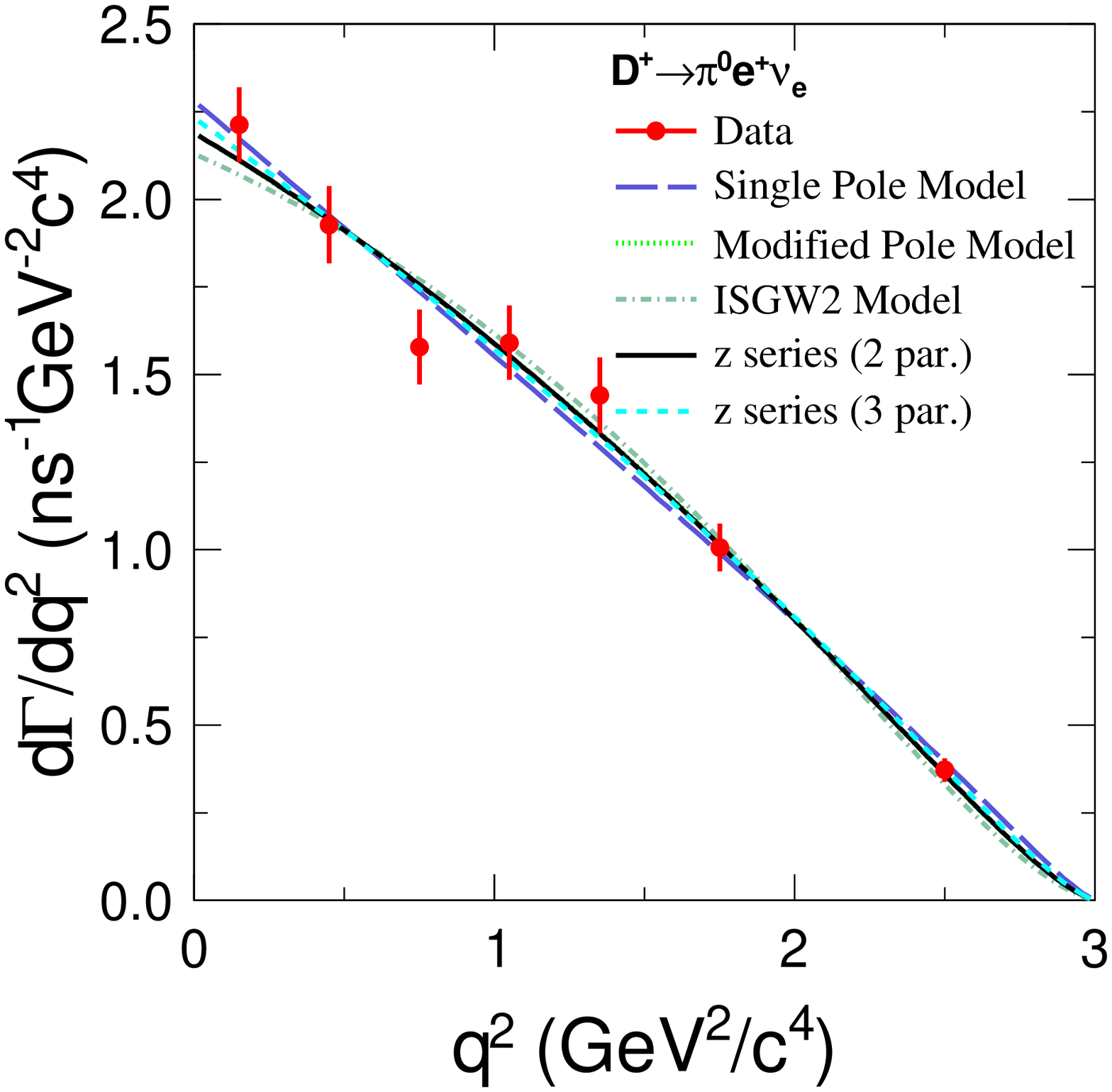} }
\end{center}
\vskip-0.20in
\caption{BESIII results~\cite{bes3:semi_Ken} for 
$D^+\ra \kbar e^+\nu$ decays (left) and 
$D^+\ra\pi^0 e^+\nu$ decays (right). The theoretical 
predictions with floated normalizations $f^+_K(0)\cdot |V^{}_{cs}|$ 
and $f^+_\pi(0)\cdot |V^{}_{cd}|$ are superimposed.}
\label{fig:bes3}
\end{figure}

BESIII has recently presented new measurements of 
$D^+\ra \kbar e^+\nu$ and $D^+\ra\pi^0 e^+\nu$ decays using hadronic 
tagging and 2.93~fb$^{-1}$ of data~\cite{bes3:semi_Ken}. The decay 
rates are measured in bins of $q^2$, as shown in Fig.~\ref{fig:bes3}.
The data points are fit to Eq.~(\ref{eqn:semileptonic}) using
several theoretical models for $f^+(q^2)$; the 
floated parameters are the normalizations 
$f_+^K(q^2\!=\!0)\cdot |V^{}_{cs}|$ and 
$f_+^\pi(q^2\!=\!0)\cdot|V^{}_{cd}|$.
Taking the form factor normalizations $f_+^K(0)$ and $f_+^\pi(0)$ 
from lattice QCD calculations~\cite{hpqcd:semi_fK,hpqcd:semi_fp},
one obtains
$|V^{}_{cs}| = 0.944\,\pm 0.005\,\pm 0.015\,\pm 0.024$ and
$|V^{}_{cd}| = 0.210\,\pm 0.004\,\pm 0.001\,\pm 0.009$, where
the third error is due to theoretical uncertainty in the 
lattice calculations. These values are consistent with 
CKM unitarity (see below).

Several other semileptonic decays have also been measured
by BESIII, although no form factor calculations for these exist:
$D^+\ra \phi(\mu^+, e^+)\nu$~\cite{bes3:semi_philn}, 
$D^+\ra \eta^{(')}\mu^+\nu$~\cite{bes3:semi_philn}, 
$D^+\ra \eta^{(')}e^+\nu$~\cite{bes3:semi_etaen}, 
$D^+\ra\kbar\mu^+\nu$~\cite{bes3:semi_Kmn}, 
and the radiative leptonic decay
$D^+\ra\gamma e^+\nu$~\cite{bes3:semi_gen}.
BESIII results for purely leptonic decays $D^+_s\ra\mu^+\nu$ 
and $D^+_s\ra\tau^+\nu$ are given in Ref.~\cite{bes3:leptonic_mn}.

\subsection{HFLAV world averages}

\begin{figure}[htb]
\begin{center}
\hbox{
\hskip-0.20in
\includegraphics[width=6.4cm]{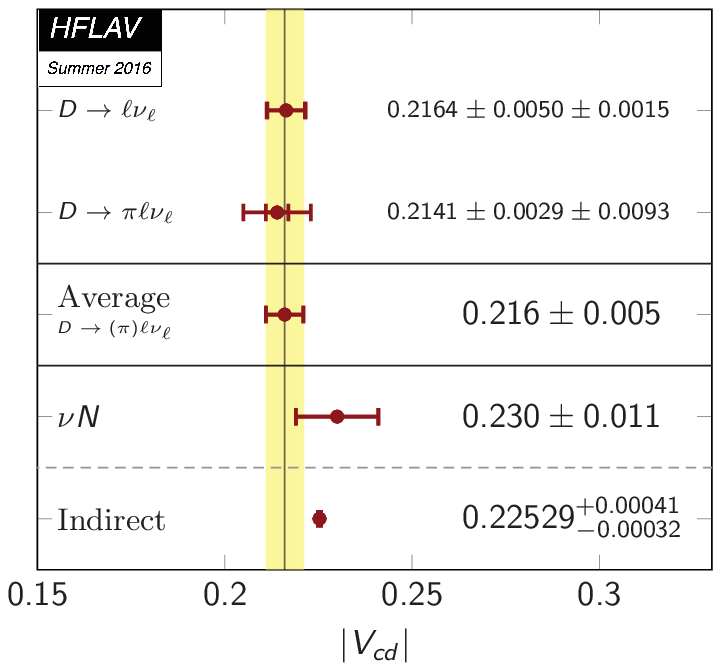}
\includegraphics[width=6.4cm]{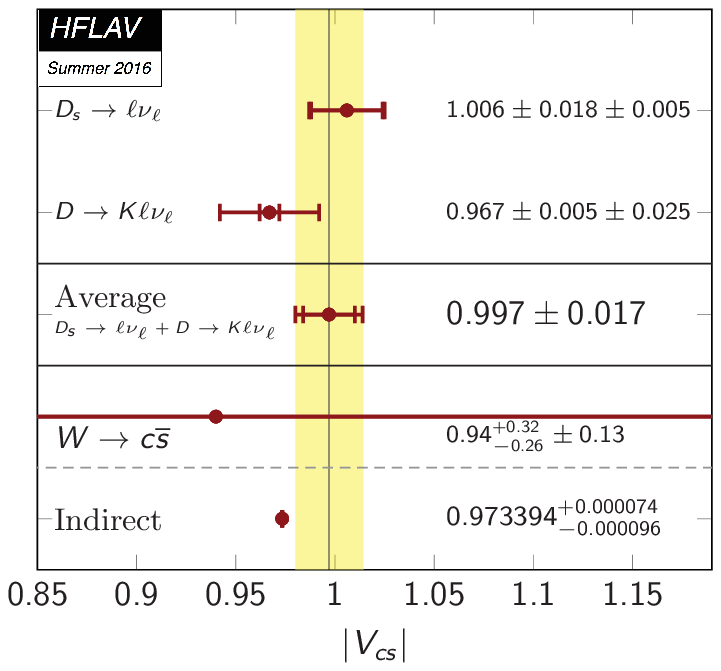} }
\end{center}
\vskip-0.30in
\caption{HFLAV world average values for $|V^{}_{cd}|$ (left) 
and $|V^{}_{cs}|$ (right), from Ref.~\cite{hflav}.}
\label{fig:hflav_leptonic}
\end{figure}

HFLAV has calculated world averages for the product 
$f^{}_D |V^{}_{cd}|$ measured using $D^+\ra\mu^+\nu$ decays, 
and for the product $f^{}_{D_s} |V^{}_{cs}|$ measured using 
$D^+_s\ra e^+\nu/\mu^+\nu/\tau^+\nu$ decays~\cite{hflav}. 
In the former case, inserting the lattice result
$f^{}_D = 212.15\,\pm 1.45$~MeV~\cite{flag:semi_fD} 
gives $|V^{}_{cd}| = 0.2164\,\pm 0.0050\pm 0.0015$,
which is consistent with the unitarity constraint
$|V^{}_{cd}| = 0.22492\,\pm 0.00050$~\cite{PDG:ckm_unitarity}. 
Averaging this result with the corresponding value from 
semileptonic $D\ra\pi\ell\nu$ decays gives a world average 
of $0.216\,\pm 0.005$, as shown in Fig.~\ref{fig:hflav_leptonic}. 
This value is also consistent with unitarity. Alternatively, 
inserting the unitarity value for $|V^{}_{cd}|$ gives 
$f^{}_D = 203.7\,\pm 4.9$~MeV, which is $1.7\sigma$ 
lower than the lattice QCD prediction.

For $D^+_s\ra\ell^+\nu$ decays, using the lattice result
$f^{}_{D_s} = 248.83\,\pm 1.27$~MeV~\cite{flag:semi_fD} gives 
$|V^{}_{cs}| = 1.006\,\pm 0.018\,\pm 0.005$, which is consistent 
with the unitarity constraint 
$|V^{}_{cs}| = 0.97351\,\pm 0.00013$~\cite{PDG:ckm_unitarity}. 
Averaging this result with the corresponding value from 
$D\ra K\ell\nu$ decays gives a world average of 
$0.997\,\pm 0.017$ (see Fig.~\ref{fig:hflav_leptonic}), which 
is also consistent with unitarity. Alternatively, inserting 
the unitarity value for $|V^{}_{cs}|$ gives 
$f^{}_{D_s} = 257.1\,\pm 4.6$~MeV, which is 
$1.7\sigma$ higher than the lattice QCD prediction.

\section{Charm baryons}

There has recently been a profusion of new measurements of charm
baryon decays. Belle, BESIII, and LHCb dominate these measurements,
and their most recent results are listed in Table~\ref{tab:charm_baryons}.

\begin{table}
\begin{center}
\begin{tabular}{|l|c|l|}
\hline
Result & Data (fb$^{-1}$) & Experiment \\
\hline
$\Lambda^+_c\ra \Sigma\pi\pi$       & 711 & Belle 2018~\cite{belle:baryons_berger} \\
$\Xi^{}_c(2930)\ra \Lambda^+_c K^-$ & 711 & Belle 2018~\cite{belle:baryons_li} \\
Excited $\Omega^{}_c$              & 980 & Belle 2018~\cite{belle:baryons_yelton3} \\
$\Omega^0_c$ hadronic decays       & 980 & Belle 2018~\cite{belle:baryons_yelton2} \\
$\Lambda^+_c\ra\phi p\pi^-\!,\,K^-\pi^+\pi^0$ & 915 & Belle 2017~\cite{belle:baryons_pal} \\
$e^+e^-\ra \Lambda^+_c \Sigma^0_c\,,\,\Xi^0_c, \Omega^0_c$ & 800 & Belle 2017~\cite{belle:baryons_niiyama} \\
Excited $\Xi^0_c\,,\,\Xi^+_c$      & 980 & Belle 2016~\cite{belle:baryons_yelton1} \\
$\Xi^{}_c(3055)\ra \Lambda D$      & 980 & Belle 2016~\cite{belle:baryons_kato} \\
$\Lambda^+_c\ra p K^+\pi^-$        & 980 & Belle 2016~\cite{belle:baryons_yang} \\
\hline
$\Lambda^+_c\ra p\mu^+\mu^-$       & 3.0 & LHCb 2017~\cite{lhcb:baryons_Lc_pmm} \\
$\Lambda^+_c\ra p K^+ K^-\!,\, p\pi^+\pi^-$\ \cpv & 3.0 & LHCb 2017~\cite{lhcb:baryons_Lc_pKK_CPV} \\
$\Lambda^+_c\ra p K^+ K^-\!,\, p\pi^+\pi^-\!,\, p\pi^- K^+$ & 1.0 & LHCb 2018~\cite{lhcb:baryons_Lc_pKK} \\
$\Xi^{++}_{cc}$                     & 1.7 & LHCb 2017~\cite{lhcb:baryons_Xcc} \\
Excited $\Omega^0_c\ra \Xi^+_c K^-$ & 3.3 & LHCb 2017~\cite{lhcb:baryons_Oc} \\
\hline
$\Lambda^+_c\ra \Xi^0 K^+,\,\Xi^0(1530)K^+$ & 0.567 & BESIII 2018~\cite{bes3:baryons_Lc_XK} \\
$\Lambda^+_c\ra \Sigma^-\pi^+\pi^+\pi^0$ & 0.567 & BESIII 2017~\cite{bes3:baryons_Lc_Sppp} \\
$\Lambda^+_c\ra p\eta ,\, p\pi^0$   & 0.567 & BESIII 2017~\cite{bes3:baryons_Lc_pe} \\
$\Lambda^+_c\ra \Lambda\mu^+\nu$    & 0.567 & BESIII 2017~\cite{bes3:baryons_Lc_Lmn} \\
$\Lambda^+_c\ra n K^0_S\,\pi^+$     & 0.567 & BESIII 2017~\cite{bes3:baryons_Lc_nKSp} \\
$\Lambda^+_c\ra pK^+ K^-\!,\,p\pi^+\pi^-$ & 0.567 & BESIII 2016~\cite{bes3:baryons_Lc_pKK} \\
$\Lambda^+_c\ra hhh$                & 0.567 & BESIII 2016~\cite{bes3:baryons_Lc_hhh} \\
$\Lambda^+_c\ra \Lambda e^+\nu$     & 0.567 & BESIII 2015~\cite{bes3:baryons_Lc_Len} \\
\hline
\end{tabular}
\end{center}
\vskip-0.10in
\caption{Recent results for charm baryon decays, from 
Belle (upper), LHCb (middle), and BESIII (bottom).}
\label{tab:charm_baryons}
\end{table}

\begin{figure}[htb]
\begin{center}
\includegraphics[width=7.2cm]{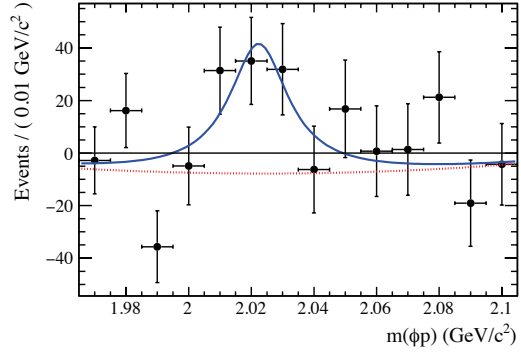}
\end{center}
\vskip-0.20in
\caption{Background-free $M(\phi p)$ invariant mass distribution 
for $D^0\ra\phi p\pi^0$ decays, from Belle~\cite{belle:baryons_pal}.}
\label{fig:phip_Bilas}
\end{figure}

An interesting result from Belle is that of a search for a
``hidden-strangeness'' pentaquark ($P^+_s$) with quark content
$s\bar{s}uud$~\cite{belle:baryons_pal}. This state would be 
analogous to the ``hidden charm'' pentaquark $P^+_c = c\bar{c}uud$ 
observed by LHCb~\cite{lhcb:baryons_penta}. For this 
analysis Belle reconstructed $\Lambda^+_c\ra \phi p\pi^0$
decays and fitted for the signal yield in bins of $M(\phi p)$
invariant mass. Plotting these yields gives a background-free
$M(\phi p)$ distribution; a peaking structure would indicate 
an intermediate $P^+_s\ra\phi p$ decay. The resulting 
distribution is shown in Fig.~\ref{fig:phip_Bilas}. 
There is an excess of events ($78\pm 28$) at 
$M^{}_{\phi p} = 2.025\pm 0.005$~GeV/$c^2$, but the 
significance is only $2.7\sigma$. The future Belle~II 
experiment~\cite{belle2}, with much higher statistics, 
should be able to clarify whether this excess is the 
first hint of a $P^+_s$ state.

\begin{figure}[htb]
\begin{center}
\includegraphics[width=6.5cm]{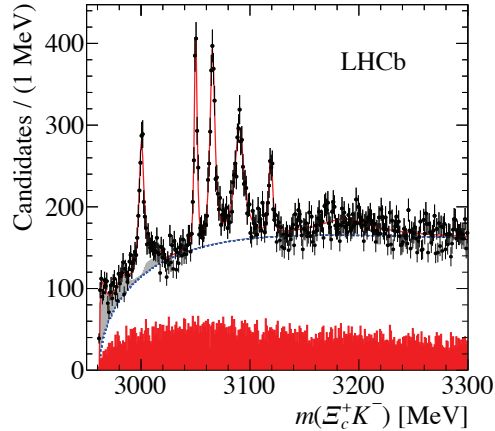}
\end{center}
\vskip-0.20in
\caption{$M(\Xi^+_c K^-)$ invariant mass distribution, 
from LHCb~\cite{lhcb:baryons_Oc}.}
\label{fig:omegac_lhcb}
\end{figure}

\begin{figure}[htb]
\begin{center}
\includegraphics[width=7.4cm]{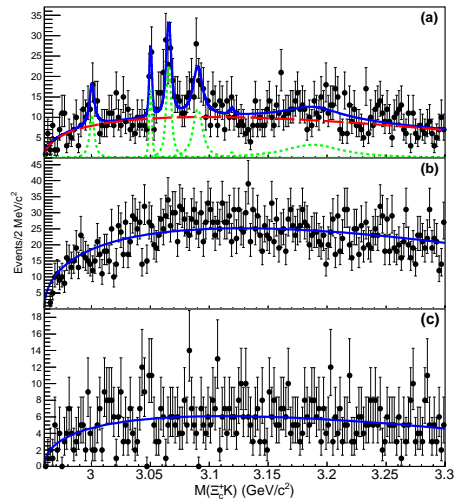}
\end{center}
\vskip-0.20in
\caption{Belle measurement~\cite{belle:baryons_yelton3} 
of excited $\Omega^*_c$ states.
Top: $M(\Xi^+_c K^-)$ invariant mass distribution.
Middle: wrong-sign $M(\Xi^+_c K^+)$ mass distribution, 
which nominally contains only background.
Bottom: $M(\Xi^+_c K^-)$ mass distribution in which the 
``$\Xi^+_c$'' is taken from the $M(p K^-\pi^+)$ sideband. 
The solid (blue) curves show the overall fit projections. }
\label{fig:omegac_belle}
\end{figure}

Another interesting result comes from both LHCb~\cite{lhcb:baryons_Oc} 
and Belle~\cite{belle:baryons_yelton2} and concerns excited $\Omega^*_c$ 
states, which have a valence quark content of~$css$. LHCb observed 
five new excited states by reconstructing $\Xi^+_c\ra pK^-\pi^+$
decays, pairing the $\Xi^+_c$ with well-identified $K^-$ tracks, 
and calculating the $M(\Xi^+_c K^-)$ invariant mass. The resulting 
distribution is shown in Fig.~\ref{fig:omegac_lhcb}. Five narrow 
peaks are observed, clearly indicating $\Omega^*_c\ra\Xi^+_c K^-$ 
decays. This result was recently confirmed by Belle (see 
Fig.~\ref{fig:omegac_belle}), although the Belle statistics 
are significantly lower and only sufficient to identify four 
of the five $\Omega^*_c$ states.

\section{Summary}

Recent world averages for $D^0$-$\dbar$ mixing and indirect 
\cpv\ parameters as calculated by HFLAV are summarized in
Table~\ref{tab:hflav_results}.
Results for searches for direct \cpv\ are summarized in 
Tables~\ref{tab:dcpv_dzero}, \ref{tab:dcpv_dplus}, and 
\ref{tab:dcpv_ds}. The most recent world averages for 
$|V^{}_{cd}|$ and $|V^{}_{cs}|$ as calculated from 
measurements of semileptonic and leptonic decays are 
plotted in Fig.~\ref{fig:hflav_leptonic}; the resulting 
values are consistent with CKM unitarity. Finally, the 
most recent results for charm baryon decays are listed 
in Table~\ref{tab:charm_baryons}. Although no statistically 
significant anomaly or ``smoking gun'' of new physics is seen, 
the precision of these results will be significantly improved 
with the analysis of LHCb Run~2 data and the large $e^+e^-$ 
dataset to be collected by Belle~II. Many new charm baryon 
measurements are expected, well beyond those listed in 
Table~\ref{tab:charm_baryons}.

\vskip0.20in

We thank the workshop organizers for hosting a productive 
meeting with excellent hospitality. The author also thanks
Andrea Contu for reviewing this manuscript.

\end{document}